\documentclass[iop]{emulateapj}

\usepackage{graphicx}
\usepackage{multirow}
\usepackage{amsmath}
\usepackage{xcolor}
\usepackage{rotating}
\usepackage{threeparttable}

\shorttitle{The HNC/HCN ratio}
\shortauthors{Graninger et al.}

\begin{document}

\title{The HNC/HCN Ratio in Star-Forming Regions}

\author{Dawn M. Graninger}
\affil{Harvard-Smithsonian Center for Astrophysics, Cambridge, MA, 02138}
\email{dgraninger@cfa.harvard.edu}

\author{Eric Herbst\altaffilmark{1}}
\affil{Department of Chemistry, University of Virginia, Charlottesville, VA, 22901}

\author{Karin I. \"{O}berg}
\affil{Harvard-Smithsonian Center for Astrophysics, Cambridge, MA, 02138}

\and
\author{Anton I. Vasyunin}
\affil{School of Physics and Astronomy, University of Leeds,  Leeds, LS2 9JT, UK}

\altaffiltext{1}{Departments of Astronomy and Physics, University of Virginia}

\begin{abstract}
HNC and HCN, typically used as dense gas tracers in molecular clouds, are a pair of isomers that have great potential as a temperature probe because of temperature dependent, isomer-specific formation and destruction pathways. Previous observations of the HNC/HCN abundance ratio show that the ratio decreases with increasing temperature, something that standard astrochemical models cannot reproduce. We have undertaken a detailed parameter study on which environmental characteristics and chemical reactions  affect the HNC/HCN ratio and can thus contribute to the observed dependence. Using existing gas and gas-grain models updated with new reactions and reaction barriers, we find that in static models the H + HNC gas-phase reaction regulates the HNC/HCN ratio under all conditions, except for very early times. We quantitively constrain the combinations of H abundance and H + HNC reaction barrier that can explain the observed HNC/HCN temperature dependence and discuss the implications in light of new quantum chemical calculations. In warm-up models, gas-grain chemistry contributes significantly to the predicted HNC/HCN ratio and understanding the dynamics of star formation is therefore key to model the HNC/HCN system.
\end{abstract}


\section{Introduction}

Isomers are prevalent in the interstellar medium, with many of the currently detected molecules having an isomeric counterpart \citep{Remijan05}. These molecules provide a unique opportunity to probe physical properties of interstellar systems if line emission ratios are regulated by a single or small set of physical characteristics.  Many isomers have a dominant species, which is typically the more stable form. For example, both HCO$^+$ and HOC$^+$ are observed in dense molecular clouds and photon dominated regions, but the ratio of HCO$^+$/HOC$^+$ is greater than 300 in these regions \citep{Smith02}. Isomer pairs are often observed in the same environments, indicating connected formation pathways and that energetic differences and selective destruction routes control the ratio. However, favoring the lower energy isomer does not hold for all ratios; the isomer pair HNC and HCN exhibits a ratio of unity at low temperatures \citep{Irvine84, Schilke92, Ungerechts97}.

The observed ratio of unity at low temperatures is counterintuitive due to the reactivity of HNC, the  metastable isomer of HCN. These isomers have a ground state energy difference of approximately 7240 K \citep{Bowman93}. Because of the relatively high abundances of both isomers and nearly identical dipole moments, implying similar excitation characteristics, this ratio has great promise as a molecular probe. The ratio has been measured in a variety of different sources, from dark clouds \citep{Schilke92} to prestellar cores \citep{Padovani11}, Titan \citep{Hebrard12}, OMC-1 \citep{Schilke92} and other galaxies and their outflows \citep{Aalto12}. A comprehensive survey designed to understand the HNC/HCN abundance ratio, and directed at OMC-1, was performed by \citet{Schilke92} using the IRAM 30m telescope and Plateau de Bure Interferometer in the 90 GHz region.  It was found that the HNC/HCN ratio is approximately 1/80 towards Orion-KL but rises to 1/5 at lower temperature regions next to Orion-KL. In the coldest OMC-1 regions, the ratio rises further to unity. The unusual temperature dependence indicates that the ratio must be kinetically controlled \citep{Herbst00}. 

Understanding the kinetics that determine the HNC/HCN ratio is key to developing HNC/HCN as a widespread tool for probing the thermal history of interstellar regions. The chemistry of the two species was studied by \citet{Schilke92} with a gas-phase astrochemical model to explain the observed HNC/HCN temperature dependence. The most important formation mechanism was found to be the dissociative recombination of HCNH$^+$, forming both HNC and HCN in a 1:1 ratio. This branching ratio has been verified experimentally by \citet{Mendes12}. The same model predicted that the temperature dependence of the HNC/HCN ratio is regulated by the destruction efficiency of HNC with H and O at different temperatures (see also \citet{PineauDesForets90}). At that time, the activation energy barriers for these reactions had not been calculated theoretically, so \citeauthor{Schilke92} set the barriers to be 200 K, which best reproduced observations of OMC-1.

Quantum chemical studies of these barriers have now been carried out and the results are used in many astrochemical networks, such as KIDA \citep{Wakelam12}. For the HNC + O reaction, a study by \citet{Lin92} determined the barrier to be approximately 1100 K. The rate coefficient of the HNC + H reaction has been calculated on a number of occasions. \citet{Talbi96} studied this reaction and determined the barrier  to be 2000 K, an order of magnitude larger than the barrier presented by \citet{Schilke92}. A more recent barrier is 1200 K (D. Talbi, private communication, 2013). Additionally, it should be noted that the backwards reaction of H + HNC (i.e., H + HCN) has a very high energy barrier of 9000 K, and is therefore not included in our study of the HNC/HNC ratio.

In addition to new barrier calculations, recent years have also seen advances in gas and gas-grain modeling of interstellar chemistry \citep{Chang12, Vasyunin13, Garrod13}. Considering the potential utility of HNC and HCN, it is therefore timely to reevaluate our understanding of what regulates this important chemical ratio. Observational advances also motivate a deeper understanding of the HNC/HCN ratio. With the near completion of the Atacama Large Millimeter/sub-millimeter Array (ALMA), its unprecedented spacial resolution and sensitivity will allow for better mapping of many interstellar environments. This is particularly exciting for the HNC/HCN ratio since temperature gradients in the less complicated low-mass star-forming regions occur on small size scales. ALMA also adds the possibility to study extragalactic sources at higher resolution. Detection of HNC and HCN is common in many extragalactic sources and is used as an AGN tracer, distinguishing Seyfert galaxies from starbursts \citep{Aalto02, Costagliola11}, however current single-dish telescopes and small interferometers are not able to resolve small-scale behavior of the HNC/HCN ratio in these sources.

To study the HNC/HCN temperature dependence anew and determine the important chemical pathways in light of recent progress, a variety of models were used to simulate the ratio. Specifically, the influence of the H + HNC barrier, initial H atom abundances, and the influence of dynamics were investigated  as the major drivers for the HNC/HCN chemistry. Models using  the most recently calculated H + HNC barrier of 1200 K will be the focus of our studies. The models used are described in Section 2, with the results and a discussion following in Sections 3 and 4.   The results are evaluated against the published HNC/HCN observations in OMC-1 \citep{Schilke92}.

\section{Modeling Methods}
The HNC/HCN ratio and the related chemistry were studied using two sets of astrochemical models, a gas phase model and a gas-grain model, both of which use time-dependent kinetics to calculate molecular abundances in different interstellar environments. The physics of the modeled systems remains static, i.e. temperature, density, and optical properties do not evolve with the chemistry. This is typically a reasonable approximation for gas-phase chemistry in dense regions, but not for grain-surface chemistry, where the physical time scales can be shorter than the chemical ones. We therefore also explore the HNC/HCN chemistry in a gas-grain warm-up model following the static models.

 The total hydrogen abundance, $n_{\rm H}$, or  density, is defined as
\begin{equation}
\label{density_def}
n_{\rm H} = n(H) + 2n(H_2),
\end{equation}
where $n(H)$ is  the abundance of hydrogen atoms and $n(H_2)$ is the abundance of H$_2$.  A range of densities, from 2 $\times$ 10$^4$ to 2 $\times$ 10$^6$ cm$^{-3}$, were investigated to determine general trends with the HNC/HCN ratio as well as to best describe the ratio in OMC-1. Other physical parameters to describe dark clouds are needed for the gas and gas-grain models as well. The visual extinction is set to 10, so that no external photons affect the chemistry, the cosmic-ray ionization rate per hydrogen atom  is 1.3 $\times$ 10$^{-17}$ s$^{-1}$, the grain radius is 0.1$\mu$m, and the grain density is 3.0 g cm$^{-3}$.  Table~\ref{elem_abund} lists three sets of elemental abundances with respect to $n_{\rm H}$ , to be discussed later in the text, in which all elements are initially in the form of atoms except for hydrogen, which is entirely molecular unless stated otherwise. The elemental abundance chosen represent an array of different values, including oxygen-rich and oxygen-poor elemental abundances to simulate depletion. To compare the calculated HNC/HCN ratio with observations the column density ratio of the $^{13}$C isotopomers was used from OMC-1 \citep{Schilke92}.

\subsection{The Gas-Phase Model Approach} \label{Nahoon}

Gas-phase astrochemical reaction networks are based on the assumption that grain-surface chemistry, desorption, and accretion play a minor role in the overall chemistry of the chemical system. Using only the gas-phase chemistry removes an error source since many gas-grain processes are poorly known or understood. We initially modeled the HNC/HCN chemistry with the public gas-phase Nahoon astrochemical model (http://kida.obs.u-bordeaux1.fr, Oct 2011 release) and the KIDA reaction network \citep{Wakelam12}, which is also publicly available.
The network incorporates anions \citep{Harada08} and high-temperature reactions \citep{Harada10}. The gas-phase abundances are solved by integrating equations of the type:
\begin{equation}
\label{gas-phase_rate}
\frac{dn_i}{dt} = \sum_{p,m}~k_{p,m}n_pn_m - \sum_{p \neq i}(k_{p,i}n_pn_i + k_{cr}[n_p - n_i] + k_{pd}[n_p - n_i]),
\end{equation}
where $i \neq p$ or m, $n_i$ is the gas-phase density of the ith species, $k_{p,m}$ and $k_{p,i}$ are gas-phase rate coefficients, $k_{cr}$, if non-zero, is the cosmic ray ionization/destruction rate for species $i$ and formation from any possible species $p$ , and $k_{pd}$ is the analogous photodissociation rate. Rate coefficient parameters for different types of reactions, only some of which have been studied, are stored in the reaction networks \citep{Wakelam12}.

To account for the formation of H$_2$, the Nahoon model incorporates two additional processes: H $\rightarrow$ H$_{\rm gr}$ and H$_{\rm gr}$ + H$_{\rm gr} \rightarrow$ H$_{2}$. The rates of these processes contain the assumption that 100\% of the atomic hydrogen adsorbed onto a dust grain, H$_{\rm gr}$, reacts to form H$_2$. To properly account for H atom desorption at higher temperatures, thermal desorption, H$_{\rm gr}$ $\rightarrow$ H, was added, using the rate equation of \citet{Hasegawa92}.  

In addition to updating the treatment of H$_2$ in the Nahoon model and reaction network, additional reactions were added in some of the model runs, as listed in Table \ref{added_rxns}. The majority of the reactions added were destruction mechanisms of HNC with C, CN, and C$_2$H, but the formation of HCN via CN + C$_2$H$_6$ was also added. The C + HNC reaction was added based on recent calculations (J.-C. Loison, private communication, 2013). The remaining reactions are temperature dependent reactions that were added for completeness.  Unless otherwise specified, the H + HNC reaction barrier was set to 1200 K (D. Talbi, private communication, 2013).

\subsection{The Gas-Grain Model}\label{Monaco}
While much of the simple chemistry in interstellar environments occurs in the gas-phase, dust grains can provide an additional route for molecule formation. Gaseous atoms and molecules are able to adsorb on dust surfaces without barriers (if the bonding is described by physisorption) and subsequently diffuse, although at 10 K, only the smaller species diffuse appreciably. If one adsorbed species finds another, they can react to produce a new, more complex molecule. Including these processes might be vital to understanding the HNC/HCN ratio since the effect of gas-grain chemistry on the ratio is unknown. The gas-grain chemistry was modeled using the MONACO model \citep{Vasyunin09} and the OSU gas-grain reaction network \citep{Garrod08}.

Equations (\ref{gas-phase_rate_gr}) and (\ref{grain_rate}) represent the general rate equations that govern gas and grain surface molecular abundances, respectively, and include the desorption and accretion terms that couple the gas to the surface processes:  
\begin{equation}
\label{gas-phase_rate_gr}
\frac{dn_i}{dt} = k^{des}_in^s_i -  k^{acc}_in_i +  \frac{dn_{i, gas}}{dt}, 
\end{equation} 
\begin{equation}
\label{grain_rate}
\frac{dn^s_i}{dt} = \sum_{p,m}~k^s_{p,m}n^s_pn^s_m - n^s_i\sum_{i \neq p} k^s_p n^s_p - k^{des}_in^s_i +  k^{acc}_in_i, 
\end{equation}
where $n_i$ is the gas-phase density of the ith species, $k_{p,m}$ and $k_p$ are gas-phase rate coefficients, $n^s_i$ is the surface density of the ith species, $k^s_{p,m}$ and $k^s_p$ are the surface rate coefficients, $k^{des}_i$ is the desorption rate coefficient (thermal and non-thermal) for the ith species, $k^{acc}_i$ is the accretion rate coefficient for the ith species, and  $\frac{dn_{i, gas}}{dt}$ is the pure gas-phase rate equation, seen in Equation 2 \citep{Hasegawa92}.
Grain surface properties, such as the dust-to-gas mass ratio and the surface site density, are used in rate coefficients to constrain the number of sites available for grain-surface processes. 
The dust-to-gas mass ratio is set to 0.01 and the surface site density, representative of olivine, is 1.5 $\times$ 10$^{15}$ sites cm$^{-2}$ \citep{Semenov03}. As described in Section 3.2.2 and listed in Table \ref{added_rxns}, the gas-phase reactions of HNC with C and O were added, as well as the grain-surface reactions of HNC with H and O to the OSU gas-grain reaction network. The gas-phase and grain-surface H + HNC reaction barrier was once again set to 1200 K (D. Talbi, private communication, 2013).

\section{Results}
The following subsections present the results for the gas-phase (Section 3.1) and gas-grain (Section 3.2) modeling. In Section 3.1, parameters in the gas-phase model are varied, focusing on physical input parameters (i.e. density, elemental abundance, and time), reaction network additions, the H + HNC reaction, and the effects of the initial H atom abundance. Section 3.2 focuses on the influence of reactive desorption and changes to the input parameters, reaction network, H + HNC barrier, and initial H atom abundances on the gas-grain modeling.

\subsection{Gas-Phase Modeling} \label{Gas_phase}
Gas-phase modeling was carried out at a range of densities (2 $\times$ 10$^4$ cm$^{-3}$ - 2 $\times$ 10$^6$ cm$^{-3}$) and temperatures (10 - 100 K). Prior to the analysis of the HNC/HCN results, the  observed abundances of CO, NH$_3$, N$_2$H$^+$, and HCO$^+$ in OMC -1 \citep{Murata90, Ungerechts97} were compared with model results. Agreement within an order of magnitude  generally occurs at times between 2 $\times$ 10$^5$  and 1 $\times$ 10$^8$ yrs  and densities between 2 $\times$ 10$^4$ cm$^{-3}$ and 2 $\times$ 10$^5$ cm$^{-3}$. One exception is the low abundance of N$_2$H$^+$, which is attributed to the lack of CO freeze-out in the model as CO is the primary gas-phase destroyer of N$_2$H$^+$. Using DCN observations and large velocity gradient modeling, OMC has been characterized by \citet{Schilke92} to have a density of 10$^5$ cm$^{-3}$ with a few higher density regions, presumably associated with OMC-KL,  of up to 4 $\times$ 10$^6$ cm$^{-3}$, similar to the densities inferred from the modeled molecules. The age is more difficult to constrain directly from observations because of uncertainties in dynamical and chemical dating of this complicated region. These constraints are taken into account when comparing models of the HNC and HCN chemistry with the observations in OMC-1. 

\begin{figure*}[htp]
\centering
\epsscale{1}
\plotone{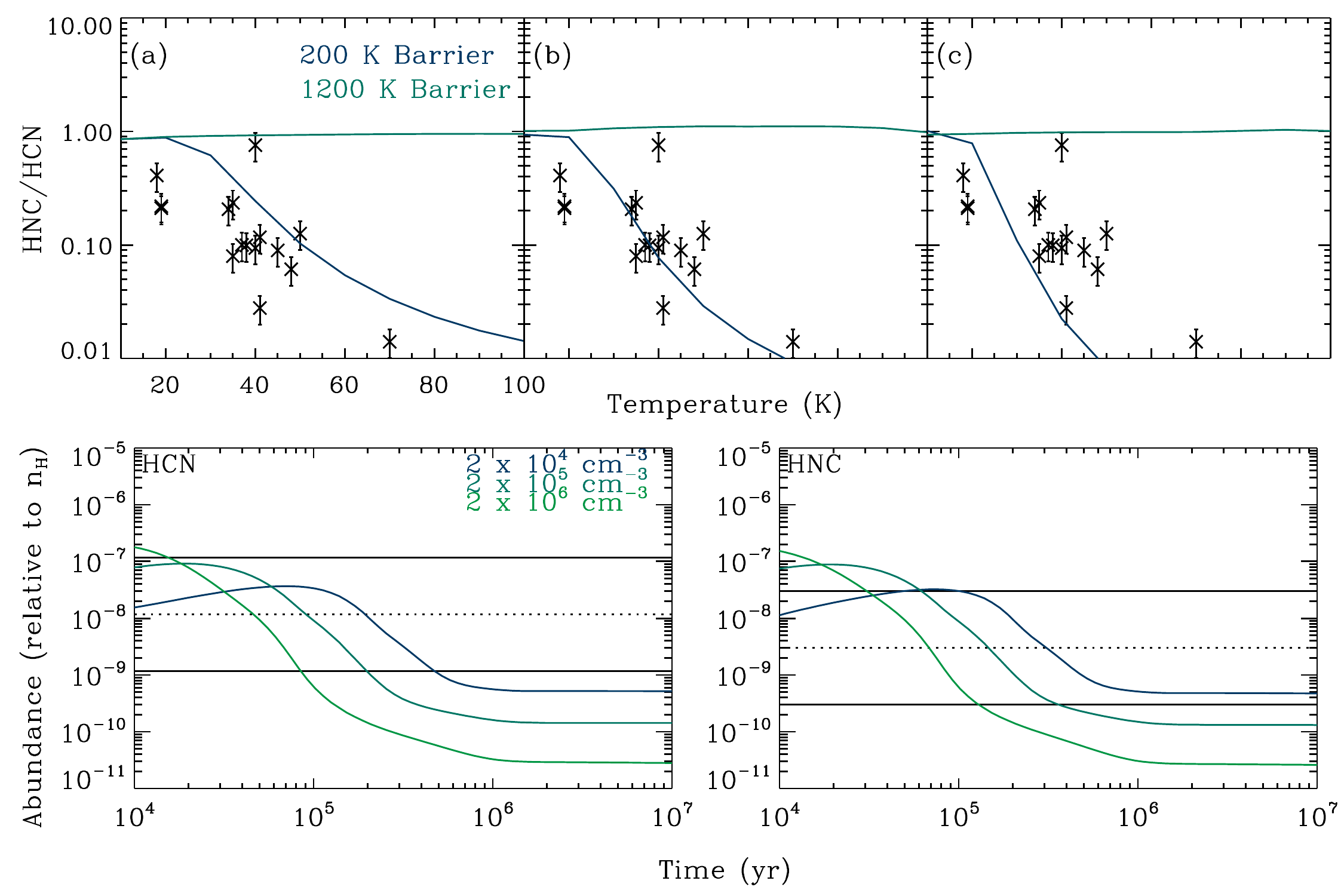}
\caption{HNC/HCN ratio and abundances of HNC and HCN for the gas-phase model. The top panels display the HNC/HCN ratio vs. temperature for densities of 2 $\times$ 10$^4$ cm$^{-3}$  (a), 2 $\times$ 10$^5$ cm$^{-3}$ (b), and 2 $\times$ 10$^6$ cm$^{-3}$ (c) at a time of 2 $\times$ 10$^5$ yrs with an HNC + H barrier of 200 and 1200 K. Here and in subsequent figures black crosses represent the observations of \citet{Schilke92} with a calibration error of 20\%. The bottom panels display the abundances of HNC and HCN as a function of time at 20 K at the same range of densities calculated using the 1200 K H + HNC barrier. The dotted lines represent the average observed abundance at 20 K. The solid black lines are $\pm$ an order of magnitude of the observed abundance. }
\label{Fig:Original_gas_mod}
\end{figure*}

The HNC and HCN abundances were calculated at different temperatures and densities, as seen in Figure \ref{Fig:Original_gas_mod}. The ratio was also calculated using the H + HNC barrier of 200 K, which was used by \citet{Schilke92}, for comparison. In the 1200 K H + HNC barrier model, the HNC/HCN ratio is close to unity at all temperatures, agreeing only with the lowest temperature observations. However, the ratio is not affected by changes in the density. The individual HNC and HCN abundances calculated using the 1200 K H + HNC barrier do depend on density;  a density of 2 $\times$ 10$^5$ cm$^{-3}$ best reproduces the observations of \citet{Schilke92}.  At this density, an agreement within an order of magnitude is achieved for both molecules at 2 $\times$ 10$^5$ yrs for all temperatures (e.g. 20 K results displayed in Figure \ref{Fig:Original_gas_mod}). Unless otherwise noted, all subsequent calculations refer to a time of $2 \times 10^{5}$ yr.

The effect of the initial elemental abundances was studied as well. The gas-phase model  includes grain reactions only for H$_2$ production, so certain elements might have larger gas-phase abundances compared with what would actually be expected due to surface accretion. With this in mind, three sets of elemental abundances, listed in Table \ref{elem_abund}, were explored. The first set of abundances are the standard low-metal oxygen-rich values, with a $\slantfrac{C}{O}$ ratio of approximately 0.4 \citep{Graedel82}. These values are known to reproduce the abundances of many observed molecules including molecular ions in dark clouds. The second set of elemental abundances incorporates C, N, and O observations of $\zeta$ Oph with heavier depletions from the gas for metallic elements \citep{Wakelam10,Cardelli93,Meyer98,Neufeld05,Graedel82}.  In this set, the  $\slantfrac{C}{O}$ ratio is 0.66 \citep{Shalabiea95}. The last set of elemental abundances incorporates a larger $\slantfrac{C}{O}$ ratio of 1.2 through a decrease in the O atom abundance, simulating H$_2$O accretion \citep{Terzieva98}. Using these three  values of elemental abundances, no change in the HNC/HCN ratio was observed at densities between 2 $\times$ 10$^4$ cm$^{-3}$ and 2 $\times$ 10$^6$ cm$^{-3}$. The low metal, oxygen-rich elemental abundances best reproduced the observed HNC and HCN abundances and are used for subsequent analyses. In summary, with the existing astrochemical networks and models,  variations in physical parameters appear to have a negligible effect on the HNC/HCN ratio. 

The mismatch between the 1200 K H + HNC barrier model and observations suggests that there are missing or underestimated HNC and HCN destruction and/or formation pathways in the existing astrochemical networks. Based on Figure \ref{Fig:Original_gas_mod}, the main problem is an over-production/under-destruction of HNC at high temperatures. Most HCN formation mechanisms also form HNC with a 1:1 branching ratio. Therefore, we focused on  selective destruction mechanisms of HNC, listed in Table~\ref{added_rxns}, which were gathered from the physical chemistry literature \citep{Hebrard12,Sims93} and J.-C. Loison  (private communication, 2013) and added to the reaction network.   The additions of these reactions do not appreciatively affect the HNC/HCN ratio at times greater than 10$^5$ yrs; at early times less than 10$^4$ yrs, the destruction of HNC with C does change the ratio. Based on these results, the H + HNC reaction remains likely to control the temperature dependence of the HNC/HCN ratio \citep{Schilke92,Talbi96, Hirota98}.

\begin{figure*}[htp]
\centering
\epsscale{1}
\plotone{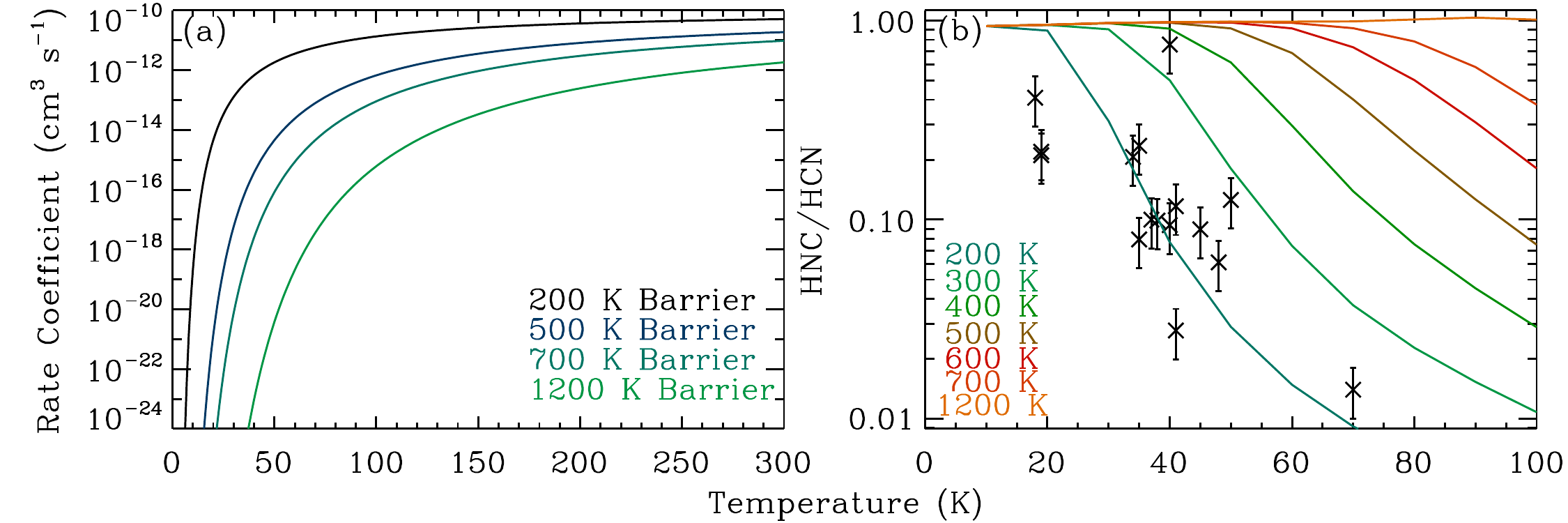}
\caption{Rate coefficients vs temperature for the H + HNC $\rightarrow$ HCN + H reaction with four different barriers (a) and calculated ratios with those barriers at 2 $\times$ 10$^5$ yrs(b). Note that in panel (a), between 20 and 40 K, there are order of magnitude differences between the commonly used 200 K barrier and the 500, 700, and 1200 K barriers.}
\label{Fig:Barrier Variation}
\label{Fig:H_HNC_rate}
\end{figure*}

If H + HNC regulates the temperature dependence of the HNC/HCN ratio, we can put an empirical constraint on its barrier via observations of the this ratio. The theoretical rate coefficient for the H + HNC reaction used in our calculations is, in units of cm$^{3}$ s$^{-1}$,
\begin{equation}
k(T) = 1 \times 10^{-10}~e^{\slantfrac{-\gamma}{T}}, 
\end{equation} 
where $\gamma$ is the barrier and T is temperature. The rate coefficient is estimated in the absence of tunneling, which tends to boost the rate coefficient somewhat at very low temperatures \citep{Talbi96}. With or without tunneling, the rate coefficient for this reaction is strongly dependent on the barrier used in the calculation within the critical temperature range (20 - 40 K) where the HNC/HCN ratio begins to decrease from near unity in OMC-1, as shown in Figure \ref{Fig:H_HNC_rate}. For example, the rate coefficient for the H + HNC reaction changes over 12 orders of magnitude without tunneling at 40 K when comparing the 200 and 1200 K barriers. With this in mind, barriers ranging between 200 and 1200 K were investigated and the observed HNC/HCN ratio was used to place a constraint on the  H + HNC barrier. At a density of 2 $\times$ 10$^5$ cm$^{-3}$, the observed HNC/HCN ratios confirm that only barriers less than 300 K are able to reproduce the high-temperature observations of the HNC/HCN ratio. 

The H atom abundance also affects the H + HNC reaction rate. The H$_2$ fractional abundance is standardly set to 0.5 at t = 0 yrs, so that all H is in H$_2$ initially and grows to a fractional abundance of approximately 10$^{-4}$. However, observations indicate that the fractional abundance of H atoms is greater than 10$^{-4}$ in dense clouds, suggesting that this value should not be initially set to zero \citep{Krco08}. Therefore, in a new set of models, the  initial H fractional abundance was varied between 1.0 $\times$ 10$^{-4}$ and 0.50 along with the barrier for the H + HNC reaction. At this density, the fractional abundance of the H atoms remains equivalent to its initial value for up to $\sim$ 1 $\times$ 10$^7$ yrs if the initial value is greater than zero.

\begin{figure}[htp]
\centering
\epsscale{1}
\plotone{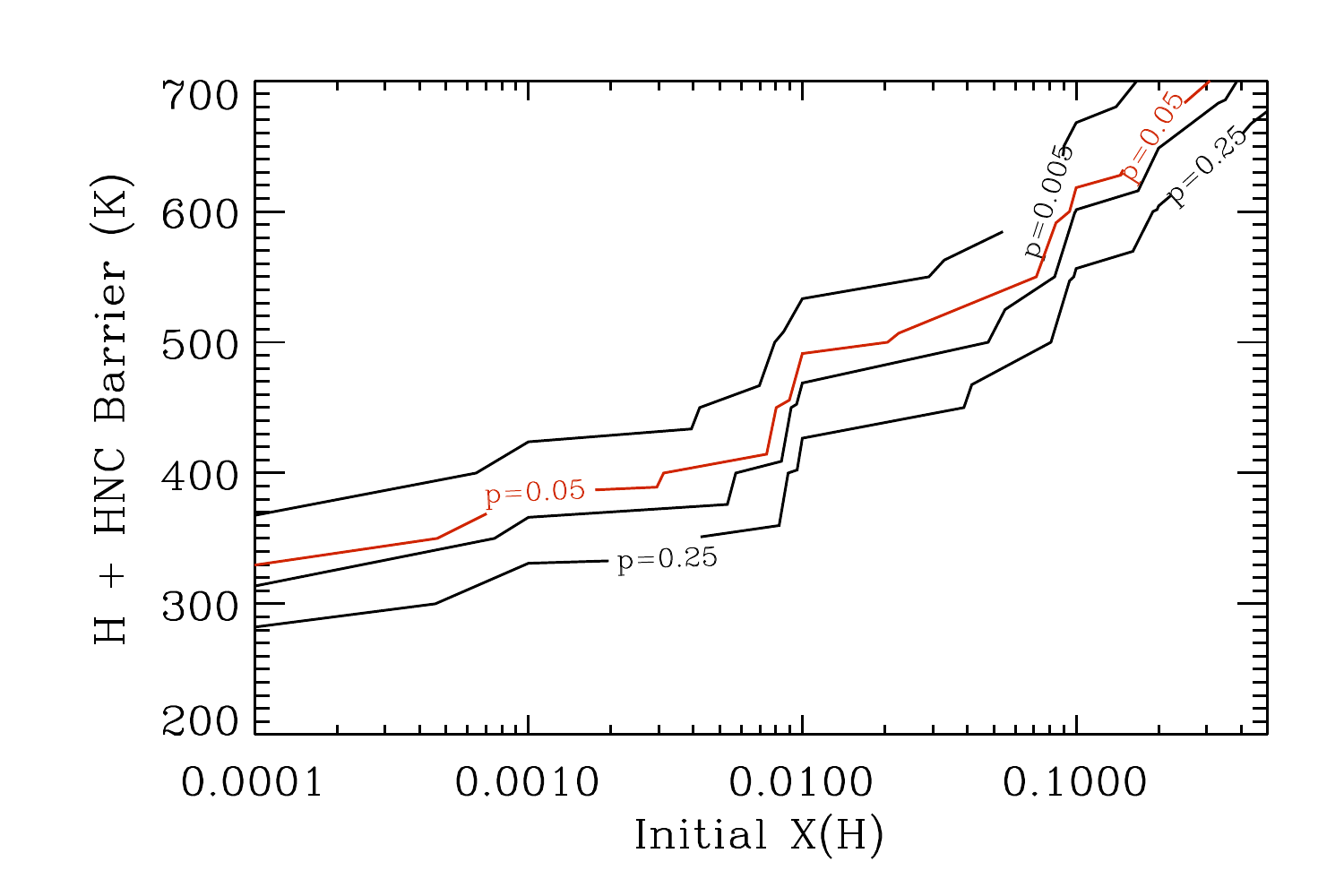}
\caption{Chi-Squared analysis of model results with the initial H abundance and H + HNC $\rightarrow$ HCN + H reaction for a density of  2 $\times$ 10$^5$ cm$^{-3}$ at  2 $\times$ 10$^5$ yrs, which best describe OMC-1. A value of $p$  less than 0.05 (red line) indicates that the model ratio statistically deviates from the observed ratio.}
\label{Fig:Chi-Square Contours}
\end{figure}

To determine how well the calculated HNC/HCN ratio reproduces the observed ratio with variations in the initial H atom abundance and H + HNC barrier, a chi-squared analysis was performed. The HNC/HCN ratios observed by \citet{Schilke92} were fit to an exponential function and then compared to the computed HNC/HCN ratios with the following equation:
\begin{equation}
\chi^2 = \sum_{i = 1}^{n}\frac{(O_i - E_i)^2}{E_i},
\end{equation}
where $O_i$ is the computed HNC/HCN ratio from the gas-phase model and $E_i$ is the expected value from our fit of the HNC/HCN  observations of \citet{Schilke92}. These $\chi^2$ values can be converted to probabilities, where a probability, $p$, of less than 5\% (or 0.05) implies a statistical deviation of the model from the observed values, in this case the HNC/HCN ratio observed by \citet{Schilke92}. Figure \ref{Fig:Chi-Square Contours} displays the results of the chi-squared analysis where the deviation threshold (p = 0.05) is indicated by a red line. Models using a barrier that lies higher than the red line for a given initial atomic H percentage statistically deviate from the observed HNC/HCN ratio. For example, if an initial fractional abundance of H is 0.01, a barrier of 600 K would deviate in a statistically significant manner whereas a barrier of 250 K could not be ruled out based on the data.  
 
In Figure \ref{Fig:Chi-Square Contours}, the H + HNC barrier can be as large as 700 K if the initial H fractional abundance is around 0.3. However, an H fractional abundance above 0.2 results in disagreements for the calculated abundance of many molecules, including HNC and HCN. The H atom abundance can be constrained further through recent observations. \citet{Krco08}, using the HI Narrow Self Absorption (HINSA) technique, determined a maximum 0.01 fractional abundance of H in L1757, a young, dark molecular cloud. Using this observational constraint, the chi-squared analysis for H atom fractional abundances less than 0.01 were further analyzed. 

\begin{figure}[hp]
\centering
\epsscale{1}
\plotone{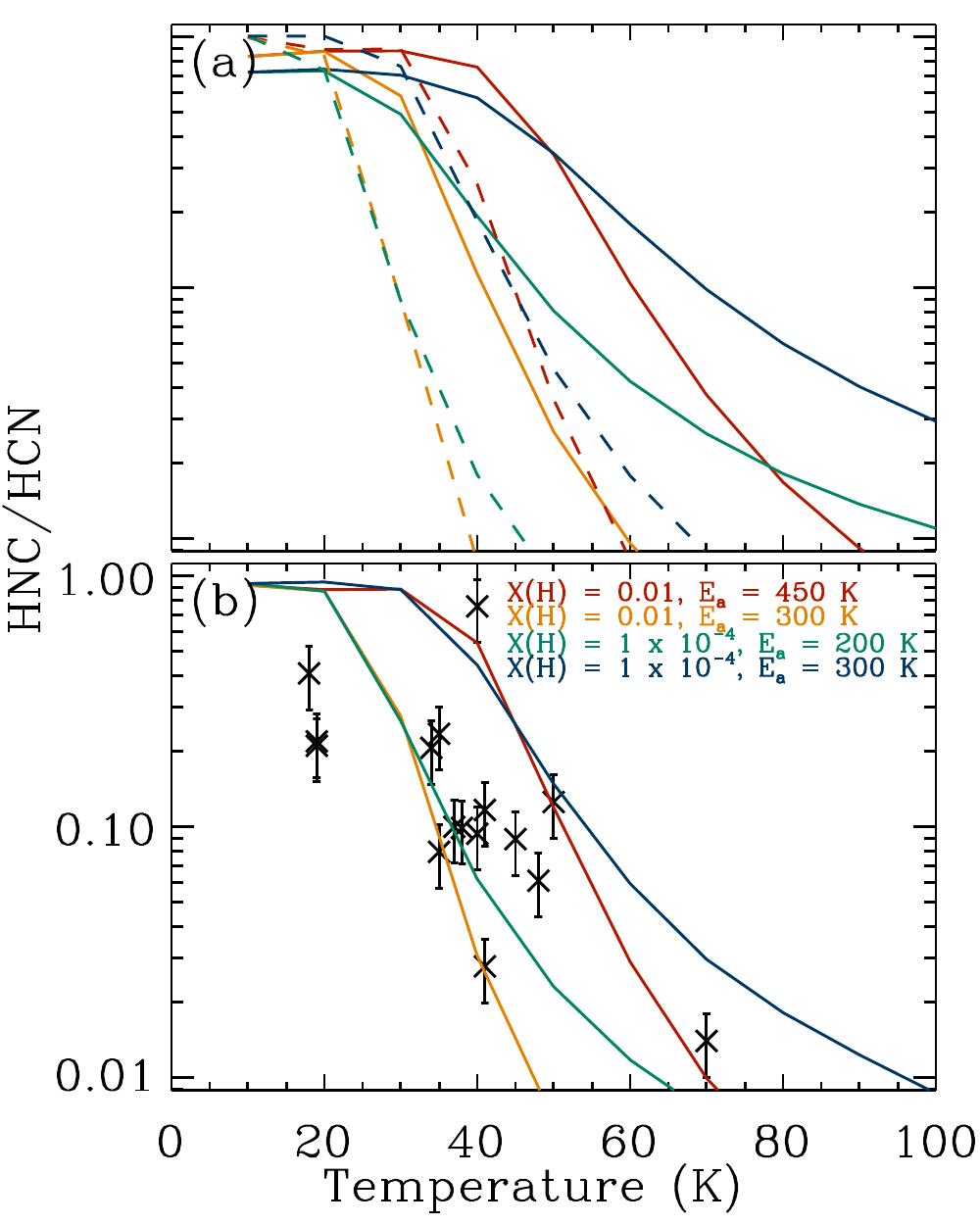}
\caption{Calculated HNC/HCN ratio  vs temperature for densities of 2 $\times$ 10$^4$ to 2 $\times$ 10$^6$ cm$^{-3}$  at 2 $\times$ 10$^5$ years. Panel (a) displays the HNC/HCN ratio at densities of 2 $\times$ 10$^4$ cm$^{-3}$ (solid lines) and 2 $\times$ 10$^6$ cm$^{-3}$ (dashed lines) and panel (b) displays the ratio at 2 $\times$ 10$^5$ cm$^{-3}$. The larger barrier for both 0.01 and 1$\times$ 10$^{-4}$ initial H fractional abundances is the maximum barrier to reproduce the observed HNC/HCN ratio and the lower barrier yields the minimum $\chi^2$.  Here and in subsequent figures, $E_a$ represents the H + HNC reaction barrier.}
\label{Fig:ratio_with_model}
\end{figure}

Figure \ref{Fig:ratio_with_model} displays the results of the best fit (minimum $\chi^2$) and the largest barrier that does not deviate significantly (p =  0.05) from the observed values of the HNC/HCN ratio at initial  fractional H abundances of 0.01 and 1 $\times$ 10$^{-4}$ for different densities. Panel (b) displays the results for a density of 2 $\times$ 10$^5$ cm$^{-3}$. This density reproduces the observations better than the lower and higher density cases, shown in panel (a). 

\begin{figure*}[htp]
\centering
\epsscale{1}
\plotone{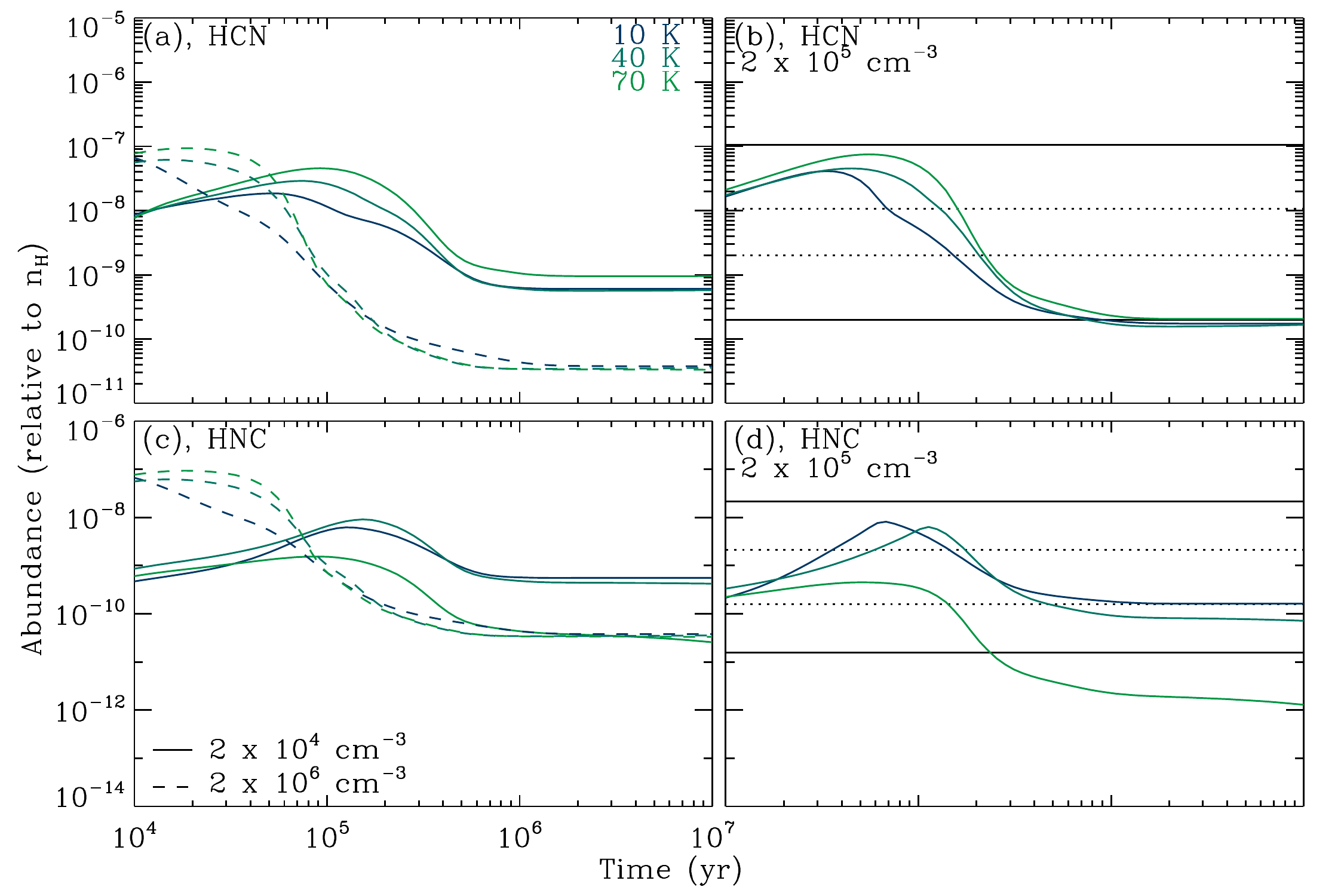}
\caption{The modeled abundances for HCN and HNC at 10, 40, and 70 K vs time with a 0.01 initial H fractional abundance and a 450 K barrier for the  H + HNC reaction, representing the highest possible barrier, at a time of 2 $\times$ 10$^5$ years. Panels (a) and (c) display the HCN and HNC abundances at densities of 2 $\times$ 10$^4$ cm$^{-3}$ (solid lines) and 2 $\times$ 10$^6$ cm$^{-3}$ (dashed lines). Panels (b) and (d) display the HCN and HNC abundances at  a density of 2 $\times$ 10$^5$ cm$^{-3}$. The dotted lines represent the max. and min. observed abundances with the solid black lines at $\pm$ an order of magnitude from these values. No comparisons are drawn to observations in panels (a) and (c) since the densities in these panels are not representative of  OMC-1.}
\label{Fig:450K_model_results}
\end{figure*}

\begin{figure*}[htp]
\centering
\epsscale{1}
\plotone{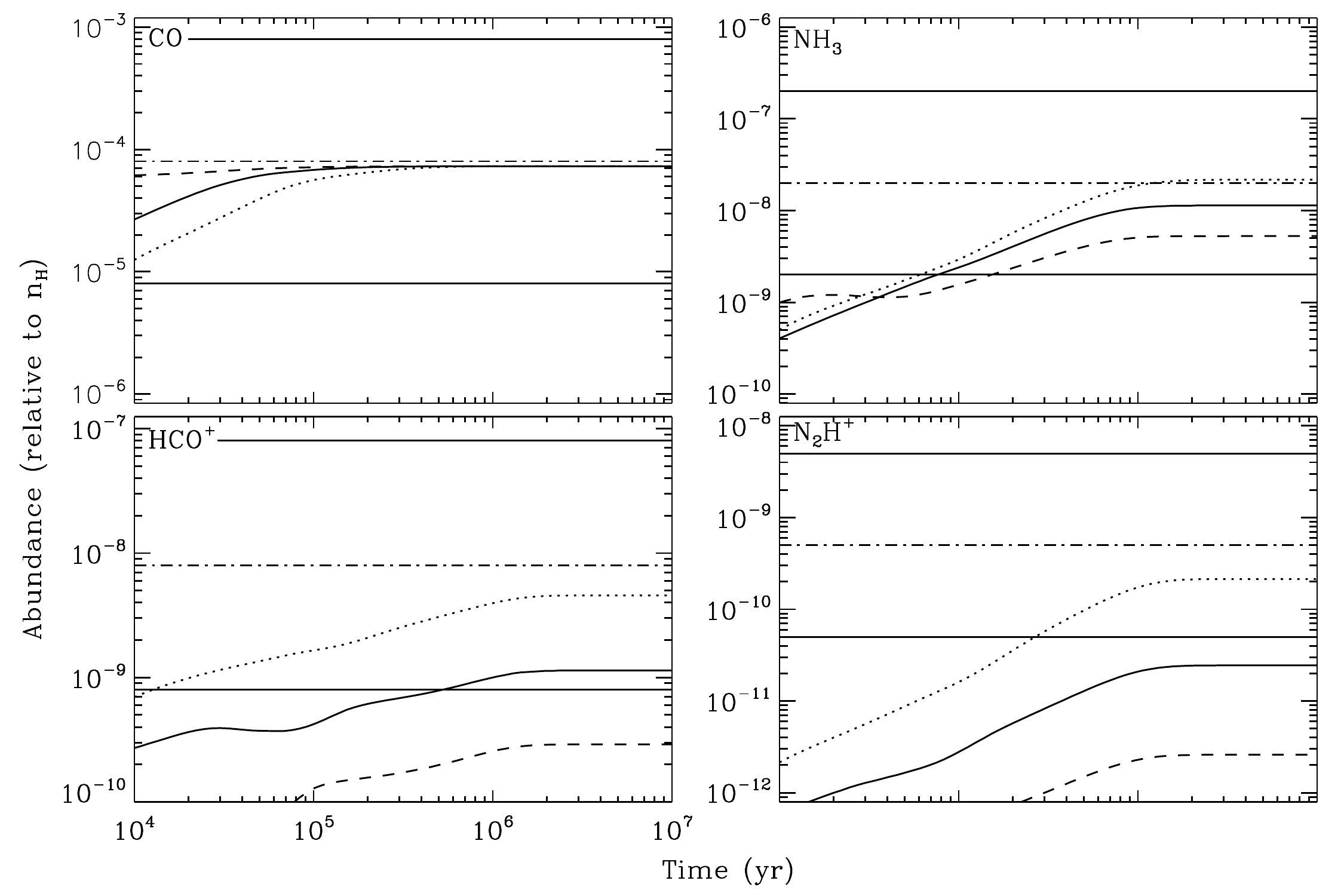}
\caption{Modeled values of common interstellar molecules with a 450 K barrier for the H + HNC reaction and 0.01 initial atomic hydrogen fractional abundance at 10 K. The dot-dashed, horizontal black lines represent the average observed abundances. The horizontal solid lines represent $\pm$ an order of magnitude from the observed abundances. The dotted and dashed lines represent densities of 2 $\times$ 10$^4$ and 2 $\times$ 10$^6$ cm$^{-3}$, respectively, while the solid line represents a density of 2 $\times$ 10$^5$ cm$^{-3}$.}
\label{Fig:Gas-phase_othermol}
\end{figure*}

Varying the H abundance may affect other aspects of the chemistry than the HNC/HCN ratio, but as shown in Figures \ref{Fig:450K_model_results} and \ref{Fig:Gas-phase_othermol}, the modeled abundances of HNC, HCN, and other molecules still agree with the observed values within approximately an order of magnitude when focusing on a density of 2 $\times$ 10$^5$ cm$^{-3}$. In summary, the reproduction of the observed HNC/ HCN ratio with current single-point gas phase models occurs only with low H + HNC barriers or a moderately high barrier ($<$ 600 K) and high initial H atom abundances. Currently, observations are reproduced satisfactorily with barriers up to 450 K if the H atom fractional abundance is 0.01. The quality of the data does not allow us to distinguish between this fit and the modeled fit with a reduced barrier and H abundance.  

\subsection{Gas-Grain Modeling}
\subsubsection{0-D, Time-Dependent Modeling}
Gas-grain modeling with the MONACO code \citep{Vasyunin09} was carried out at a range of densities (2 $\times$ 10$^4$ cm$^{-3}$ - 2 $\times$ 10$^6$ cm$^{-3}$), temperatures (10 - 100 K), and a 1\% reactive desorption efficiency to check the overall validity of this chemical model. Again, the results were compared with observations of OMC-1 \citep{Murata90, Ungerechts97} and the best agreement with observed common interstellar molecules, including HNC and HCN, occurred at times between 1 $\times$ 10$^5$ and 2 $\times$ 10$^8$ yrs  and densities of 2 $\times$ 10$^4$ cm$^{-3}$ and 2 $\times$ 10$^5$ cm$^{-3}$. The temperature dependence of the observed HNC/HCN ratio was not reproduced with the standard network and reaction barriers, so explorations of the reactive desorption efficiency, physical parameters, and reaction networks were carried out to elucidate the effect of grain chemistry  on the HNC/HCN ratio. 

\begin{figure}[htp]
\centering
\epsscale{1}
\plotone{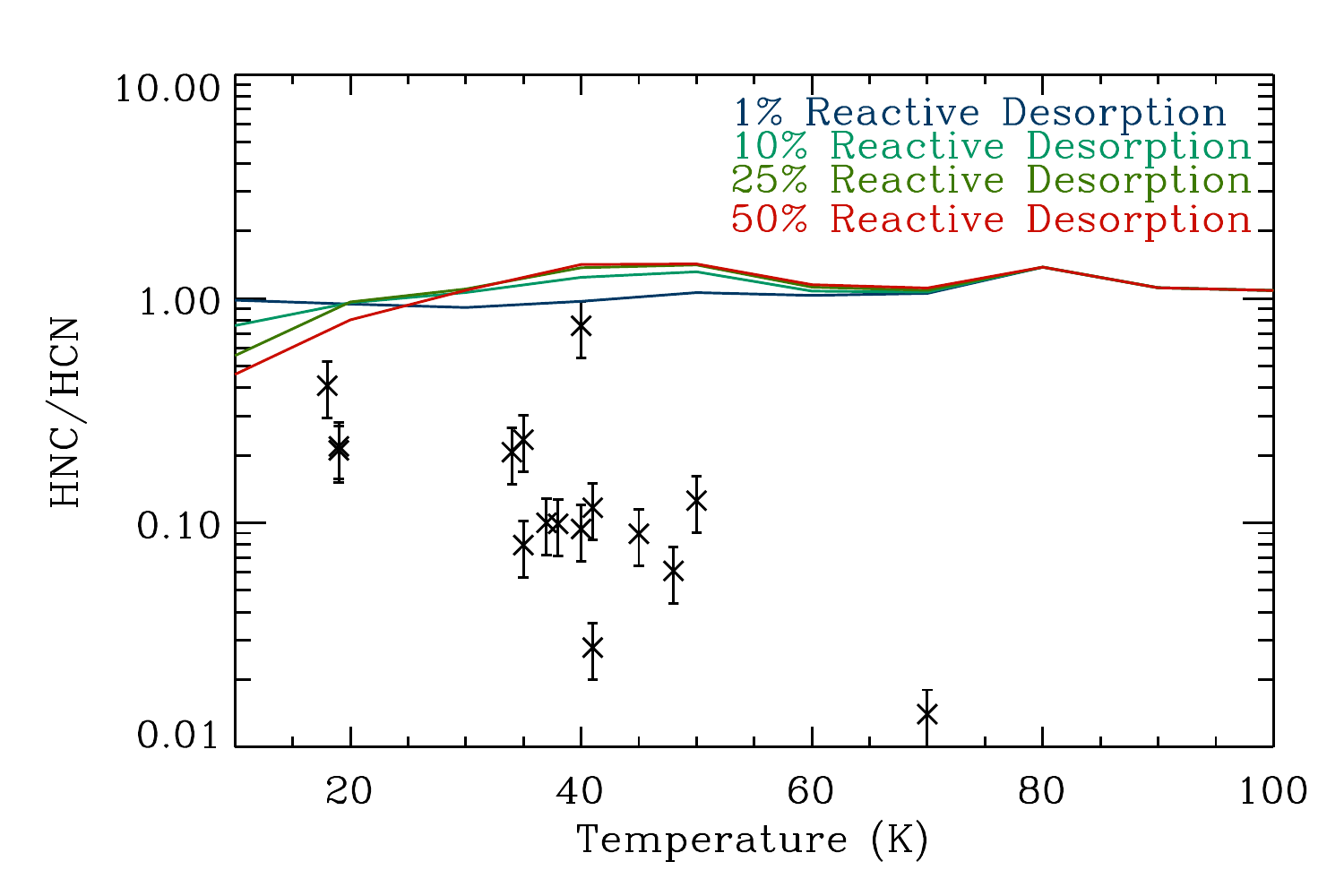}
\caption{Modeled gas-phase HNC/HCN ratio vs temperature using the MONACO model for four different efficiences of reactive desorption. The barrier for H + HNC $\rightarrow$ HCN + H is set to 1200 K and there are no initial hydrogen atoms present at a density of 2 $\times$ 10$^5$ cm$^{-3}$.}
\label{Fig:Des_reactive}
\end{figure}

Accretion and desorption are the primary processes that couple the gas-phase and grain surface chemistry. Reactive desorption is one of the lesser understood processes, having an unknown efficiency. It occurs when a fraction of the products of a grain-surface reaction do not remain on the surface, but are immediately ejected into the gas-phase via local energy transfers into the reaction product-grain bond. In Figure \ref{Fig:Des_reactive}, the reactive desorption efficiency for all grain surface reactions is varied between 1 and 50 \% at a density of 2 $\times$ 10$^5$ cm$^{-3}$ to elucidate how important grain surface chemistry is in setting the HNC/HCN gas-phase ratio. No significant change in the HNC/HCN ratio nor the HNC and HCN fractional abundances is observed when the reactive desorption efficiency is varied from 1\% to 50\%, suggesting that the grains do not play a direct role in the HNC/HCN ratio chemistry. Small changes are observed between temperatures of 30 and 60 K, but the effect is less than a factor of two.  Since no significant effect was observed in the ratio as a function of the reactive desorption efficiency at temperatures greater than 20 K, the efficiency was set to  10\% for the remainder of the paper \citep{Vasyunin13}.

\begin{figure}[htp]
\centering
\epsscale{1}
\plotone{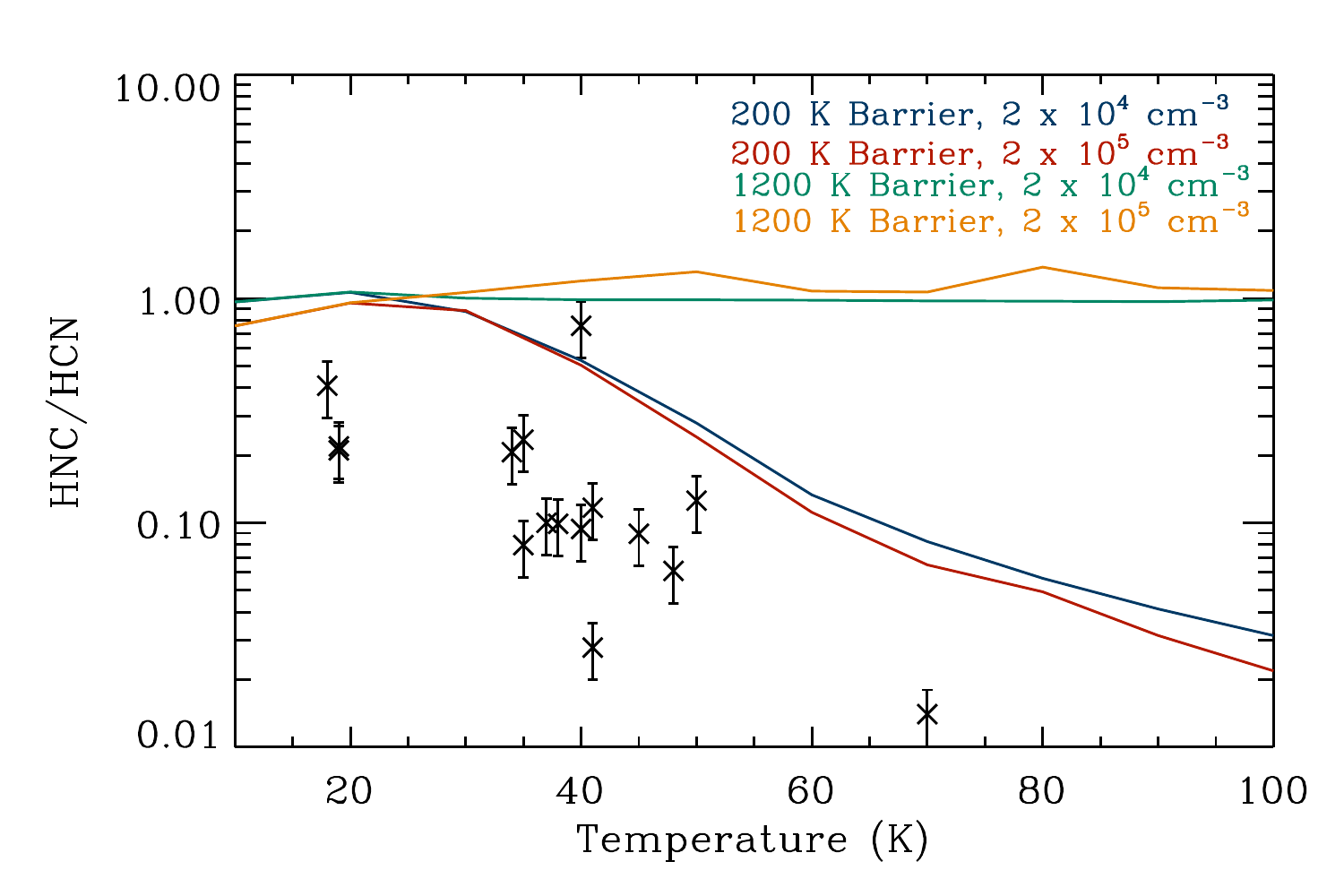}\\
\caption{Modeled HNC/HCN ratio vs temperature with no initial atom hydrogen and different H + HNC reaction barriers using the gas-grain model MONACO at 2 $\times$ 10$^5$ yrs, before the addition of any new reactions.}
\label{Fig:Gas-grain_model-ratio}
\end{figure}

Furthermore, the density, as shown in Figure \ref{Fig:Gas-grain_model-ratio}, and elemental abundances were found to have no effect on the HNC/HCN ratio, consistent the gas-phase model results. After this initial analysis, the reaction network was updated with reactions listed in Table \ref{added_rxns}. The reactions added to the network were primarily grain-surface equivalents of important reactions in the gas-phase. Additionally, the HNC + O reaction was added  along with its grain surface analog. The grain surface reactions in Table \ref{added_rxns} have the same activation barrier as the gas-phase reactions as an initial approximation. This too did not affect the HNC/HCN ratio significantly.

\begin{figure}[htp]
\centering
\epsscale{1}
\plotone{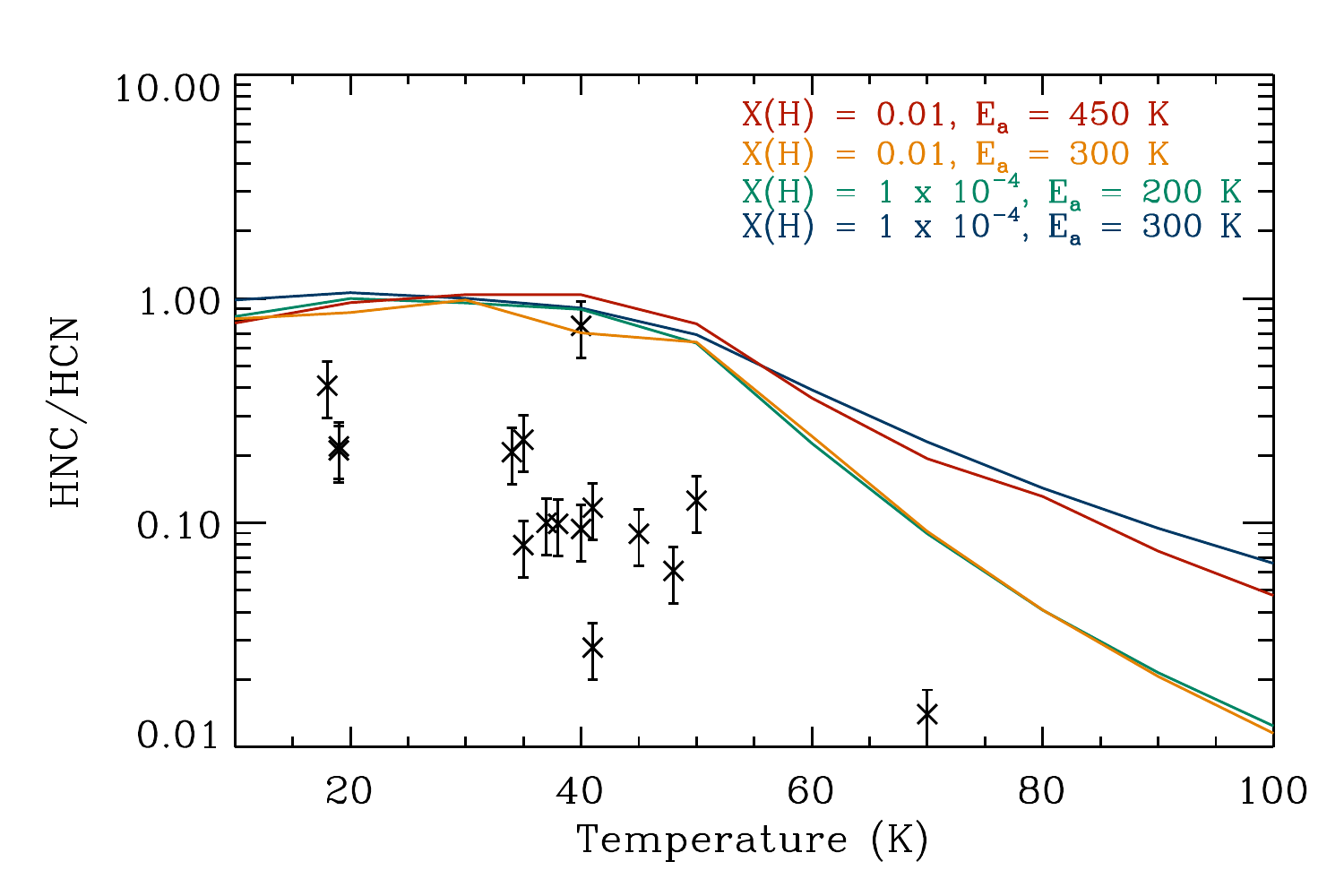}\\
\caption{Modeled HNC/HCN ratio vs temperature with initial H abundances and different H + HNC reaction barriers using the updated OSU chemical network with the gas-grain model MONACO at a density of 2 $\times$ 10$^5$ cm$^{-3}$ at 2 $\times$ 10$^5$ yrs. The initial hydrogen abundances and barriers were selected based on the gas-phase results displayed in Figure~\ref{Fig:Chi-Square Contours}.}
\label{Fig:Gas-grain_model-ratio_withH}
\end{figure}

The effects of a non-zero initial H and lowering of the H + HNC barrier were investigated to see if the gas-grain model was able to reproduce the gas-phase model results. While the full parameter space evaluation was not performed, initial H atom fractional abundances of 0.01 and 1 $\times$ 10$^{-4}$ were implemented with barriers of 300 K and 450 K for $X(H)$ = 0.01, and 200~K and 300 K for $X(H) = 1 \times$ 10$^{-4}$,  as displayed in Figure \ref{Fig:Gas-grain_model-ratio_withH}, where $X(H)$ is the fractional abundance of H atoms relative to the total proton density, or $\frac{n(H)}{n_H}$. While not identical the gas-grain and gas-phase models result in similar HNC/HCN ratios as a function of temperature. This supports the gas-phase modeling results, that the H atom abundance and H + HNC barrier together control the temperature dependence of the HNC/HCN ratio in static models. 

\subsubsection{Warm-up Models}
All previous models discussed are static 0-D models; here we discuss the results of  warm-up models with the gas-grain network where only the temperature changes as a function of time. This type of model simulates a potentially more realistic scenario where all material started out cold, followed by warm-up due to nearby star formation and is therefore important to obtain a complete understanding of which processes that can affect the temperature dependence of the HNC/HCN chemistry. Three warm-up models were run between 10 - 100 K with different, constant heating rates, resulting in warm-up times of 5 $\times$ 10$^{4}$, 2 $\times$ 10$^{5}$, and 1 $\times$ 10$^{6}$ yrs.   The shortest time refers to high-mass star formation, and the longest time to low-mass star formation \citep{Garrod06}. A fourth warm-up model with a very long warm up time of  5 $\times$ 10$^6$ yrs was also used. The models were run using H + HNC barriers of 200 K and 1200 K with no initial H fractional abundance and an H + HNC barrier of 450 K with $X(\rm H) = 0.01$ for each of the warm-up time scales.

\begin{figure*}[htp]
\centering
\epsscale{1}
\plotone{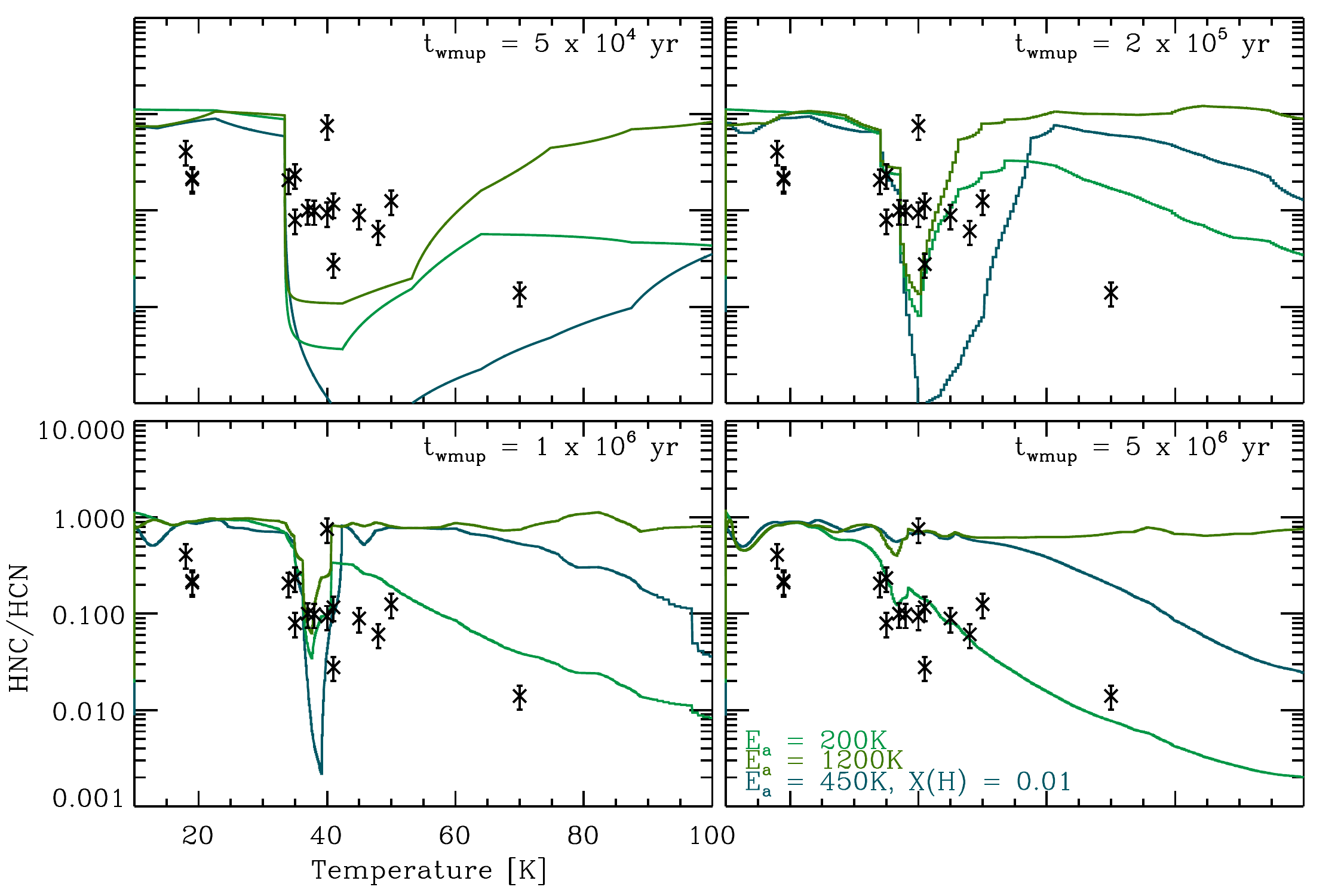}\\
\caption{Modeled HNC/HCN ratio vs temperature with initial H abundances and different H + HNC reaction barriers using the updated OSU chemical network with the gas-grain model MONACO at a density of 2 $\times$ 10$^5$ cm$^{-3}$ and a range of warm-up time scales, t$_{wmup}$. Unless specified, all initial hydrogen is in the form of H$_2$; i.e. X(H) = 0.}
\label{warm-up}
\end{figure*}

The temperature dependence of the HNC/HCN ratio for the warm-up models is displayed in Figure $\ref{warm-up}$. In our discussion, we divide the warm-up into three temperature ranges.  The first starts at 10 K and ends around $\approx 35$~K, the second encompasses a large dip in the HNC/HCN abundance ratio for all but the longest warm-up model in the vicinity of 35 - 40 K, and the last runs from 40 K to 100 K, the final temperature.    In all simulations, there is little temperature dependence in the first range  and the HNC/HCN ratio remains near unity, which is somewhat larger than the few observed values in this region. The sharp dip at 35 - 40 K reduces the HNC/HCN ratio to or below the observed values. Although not displayed in Figure $\ref{warm-up}$, the HNC and HCN abundances are also in agreement with the observed fractional abundances in all simulations  \citep{Schilke92}. The temperature dependence of the HNC/HCN in the last range of temperatures differs strongly depending upon the warm-up time and the other parameters.  For the shortest warm-up time, the HNC/HCN ratio gradually increases again, while for the two intermediate warm-up times, there is a sharper increase, especially for $1 \times 10^{6}$ yr, back to a value near unity, followed at higher temperatures by little temperature dependence for models with the 1200 K barrier, and gradual decreases for the lower barrier models.  For the longest warm-up time, there is only a small feature at 35 - 40 K, and a gradual decrease in the HNC/HCN ratio except for the high barrier model. 

Of the warm-up models, the slow warm-up/low-barrier model, which unsurprisingly simulates the static models quite well, results in the best fit to the observations. In contrast to the static models, warm-up models with higher HNC + H barriers cannot be ruled out, however, which implies that grain surface chemistry can be very important for the HNC/HCN chemistry in regions that are experiencing warm-up on short time scales.

\section{Discussion}   

The chemistry of HCN and its higher energy isomer HNC is reproduced quite well in cold dense interstellar sources by gas-phase ion-molecule chemistry.  Both the individual abundances and the abundance ratio of near unity are fit by a chemistry in which the formation of HNC and HCN is primarily via the dissociative recombination of HCNH$^+$, initially predicted by \citet{Herbst73}, while reactions with H$_3^+$, H$_3$O$^+$, and HCO$^+$ are the dominant molecular destruction mechanisms.  For sources at higher temperature, the HNC/HCN abundance ratio decreases rapidly, an unusual occurrence since one would expect the higher energy isomer to increase in abundance.  Indeed, the temperature dependence of the ratio implies that the ratio is governed by kinetic rather than thermodynamic considerations.  The temperature dependence of the HNC/HCN abundance ratio has been studied most carefully in the source OMC-1 \citep{Schilke92}, and we have endeavored to simulate this temperature dependence with up-to-date gas-phase and gas-grain chemical networks.  

\begin{figure}[hp]
  \centering
    \includegraphics[width = 0.5\textwidth]{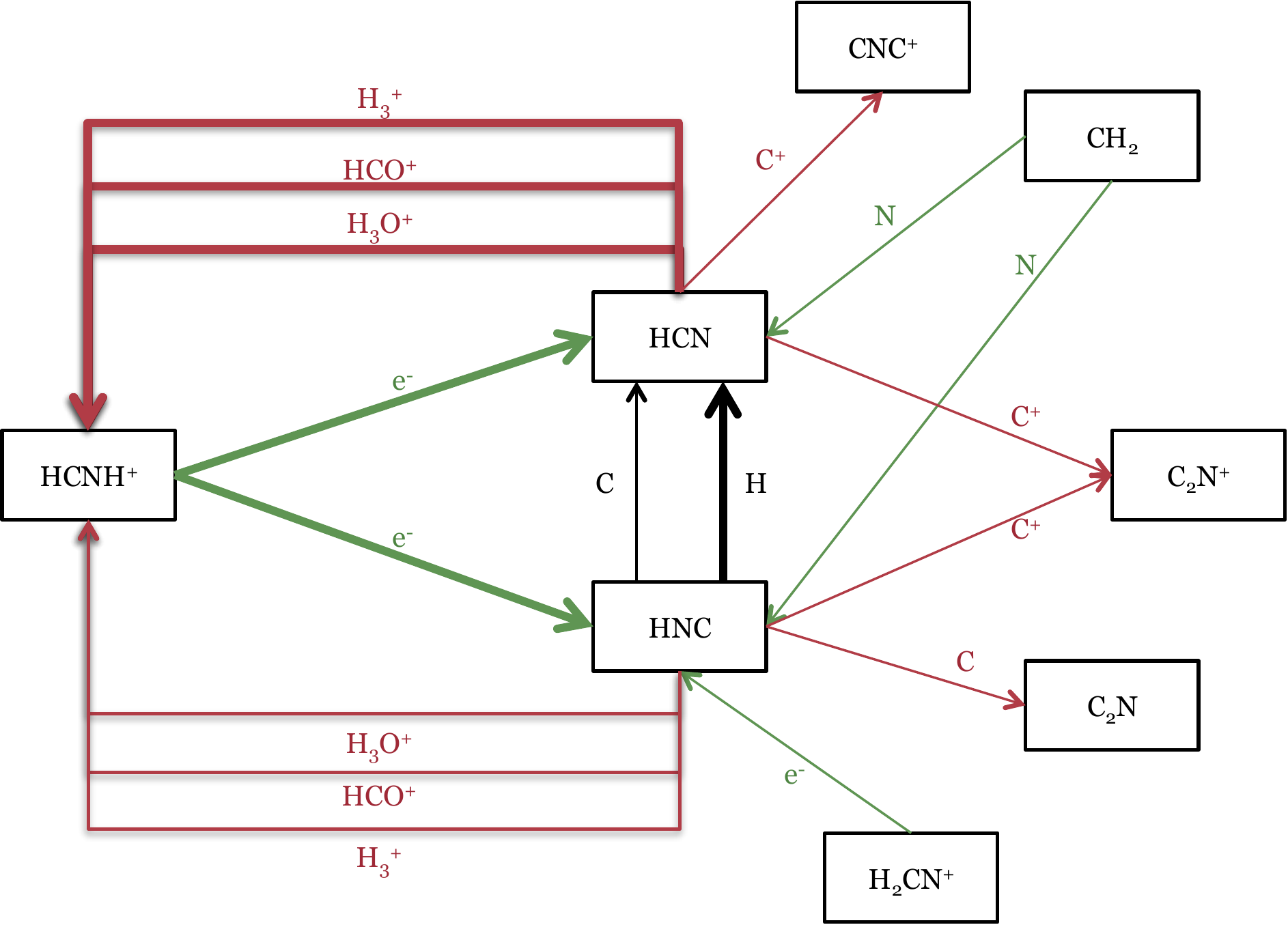}
  \caption{HNC and HCN reaction pathways for 10 to 100 K at 2 $\times$ 10$^5$ yrs. Green indicates formation, red indicates destruction, and black represents both formation and destruction with bold lines drawing attention to the most important reactions.}
    \label{pathway_gas}
\end{figure}

But what reaction or reactions can lead to the unusual HNC/HCN temperature dependence?  The gas-phase formation and destruction pathways for  the chemistry of the HNC/HCN ratio are shown in Figure \ref{pathway_gas} for chemical times up to $ 2 \times 10^{5}$ yr.   In addition to the low temperature ion-molecule processes, there are a number of destruction mechanisms for HNC involving neutral reactant partners. The only reaction that appears to be of importance in reducing the HNC/HCN ratio at times representative of dense clouds in static systems is the reaction between HNC and atomic H, which is known by a number of quantum chemical calculations to have an activation energy barrier, although the different calculations yield a wide divergence of results between the earliest calculation \citep{Talbi96} and more modern ones (700 K, 1200 K) (D. Talbi, private communication, 2013).   Even with the lowest calculated barrier (700 K), our gas-phase model calculations fail to reproduce the extent of the diminution of the HNC/HCN ratio with increasing temperature unless it is assumed that a much larger abundance of atomic hydrogen is present than in standard ion-molecule calculations.  Indeed, a chi-squared analysis shows that for a density of $ 2 \times 10^{5}$ cm$^{-3}$ and a time scale of $ 2 \times 10^{5}$ yr, the maximum barrier consistent in a statistical sense with the HNC/HCN data is 600 K, at a rather high H atom initial fractional abundance of 0.1.    If a more realistic upper limit for the fractional abundance of H of 10$^{-2}$ is utilized, we obtain a maximum barrier of near 500 K. With a more standard theoretical fractional abundance of 10$^{-4}$, this maximum barrier is slightly higher than 300 K.   None of these values are even close to the most recently calculated barrier of 1200 K.   We have not been able to improve upon these results by variation of parameters such as the elemental  abundances, the lifetime of the source,  and its density nor have we been able to do better with a full-fledged gas-grain chemical simulation at a constant temperature.

The inclusion of a warm-up in the gas-grain chemical simulations provides some information on how dynamics affect the system. Depending on the warm-up time-scale, different and complicated temperature dependences for the HNC/HCN ratio  are calculated, due, in part, to the enhanced importance of indirect gas-grain interactions, specifically the influence of formaldehyde. H$_2$CO forms on grain-surfaces via the hydrogenation of CO. Once H$_2$CO has desorbed, it reacts with HCNH$^+$ to form HCN, a reaction that was studied experimentally by \citet{Freeman78}. The analogous reaction to form HNC is quite endothermic, and does not occur. In our chemical simulations, the reaction becomes important between 30 - 40 K, explaining the dramatic decrease in the HNC/HCN by an enhancement in the HCN abundance. Additionally, a second, direct gas-grain process occurs in this temperature range through the desorption of HCN and HNC from the grain surface. The contribution to the 35 K dip arises  because of the large abundance of HCN on the grain surface compared with HNC.

Following this initial decrease around 35 K, the HNC/HCN ratio increases in all warm-up models; however, the rate of the increase varies depending on the warm-up time scale. This is again primarily due to the H$_2$CO reaction and time effects. With the shortest warm-up  time scale investigated of 5 $\times$ 10$^4$ yrs, very little time is spent at each temperature. Therefore, the chemistry that occurs at higher temperatures is still influenced by the lower temperature chemistry. However, as the warm-up stages get longer, the amount of time that each temperature has experienced increases, resulting in the lower temperature chemistry proceeding further before the onset of the higher temperature chemistry. As the warm-up timescales for each temperature increase, they approach the time scales used for the static models and the dip becomes negligible, analogous to the static model results. The chemistry of the HNC/HCN ratio after the dip is identical to that determined by the important reactions found in the static models,  as displayed in Figure \ref{pathway_gas}.

With all of the results from the gas-phase and gas-grain models, we can infer from the inability to reproduce the observed temperature dependence of the HNC/HCN ratio using static models that one or more of the following inferences must be true:  

\begin{enumerate}

\item  the H + HNC barrier is lower than the most recent calculated value of 1200 K.
\item the reaction networks are not complete. 
\item the observations are heavily influenced by short dynamic timescales and therefore gas-grain chemistry, consistent with the complicated nature of the source. 
\end{enumerate}

Through lowering the H + HNC barrier from the calculated value of 1200 K, this reaction becomes an increasingly efficient destruction mechanism. Differences between the empirical and theoretical barrier values could elucidate whether other processes, such as tunneling, are occurring in interstellar environments. The barrier discrepancies also suggest that we might need to update or are missing some important reactions for this chemistry. For example, the HNC + O  reaction was found to be important for the HNC/HCN ratio by \citet{Schilke92}, but was last calculated by \citet{Lin92}. Since this time, quantum chemical methods used to determine barriers have advanced so that a recalculation of the barrier or a laboratory study of this reaction might be worthwhile.

The inferences obtained from static models may not represent the situation satisfactorily.  It is also plausible that the dynamics of the observed system can produce variations of the observed HNC/HCN ratio. Indeed, with astrochemical warm-up models, the effect of warm up the chemistry on the HNC/HCN ratio is quite evident, although strongly dependent on the rate of the warm-up. The dynamic models indicate that reasonable models of the HNC/HCN dependence on temperature with barriers higher than 200 K  might exist.  However, these models only incorporate the warm-up phase of a collapse, omitting any variations in the density. Further exploration of this effect on the chemistry of the HNC/HCN ratio is necessary to fully understand observed results if a system is known to be dynamic.

The comparison between the observations of \citet{Schilke92} and our simulations for the HNC/HCN abundance ratio in OMC-1 has also been affected by new calculations by \citet{Sarrasin10} of rotationally inelastic collisions involving HCN and HNC with He.  These collisional calculations indicate that HNC/HCN abundance ratios greater than unity in cold cores inferred from prior collisional rates \citep{Green74} are likely to be too high and should be reduced to unity for sub-critical gas densities, close to our calculated values at 10 K.  For the HNC/HCN results of \citet{Schilke92}, which even at low temperatures (20 K) are less than unity, the new collisional results are likely to push the ratio somewhat lower, although the density is closer to critical, and slightly increase the disagreement between theory and observation.  The agreement could be improved by  an additional low temperature destruction mechanism for HNC and might be indicative of tunneling under an activation energy barrier, a process not currently included in the gas-phase portion of our models.

Finally, while our model to explain the HNC/HCN ratio in OMC-1 determines the important chemical parameters for this ratio, the region that was studied by \citet{Schilke92} might not be the ideal source for comparison. In addition to our use of  older observations which incorporate now out of date collisional rates, as mentioned above, OMC-1 and the region around OMC-KL are not quiescent, with an energetic outflow,  which might cause shocks \citep{Allen93}. Although, through combining multiple studies of other clouds, a similar temperature dependence of the HNC/HCN ratio is still observed (e.g.\citet{Ungerechts97, Bergin97, Pratap97, Liszt01, Jorgensen04, Padovani11}), this is also non-ideal. New observations of the HNC/HCN ratio in more quiescent sources with resolvable and known temperature structures, such as protostars and protoplanetary disks, would be ideal laboratories to develop the HNC/HCN ratio as a robust temperature probe for a variety of different interstellar regions.

\acknowledgements
E. H. wishes to acknowledge the support of the National Science Foundation for his astrochemistry program. He also acknowledges support from the NASA Exobiology and Evolutionary Biology program through a subcontract from Rensselaer Polytechnic Institute. A. Vasyunin acknowledges the financial support of the European Research Council (ERC; project PALs 320620)

\begin{table}[hp]
\begin{threeparttable}
\begin{center}
\caption{Initial  Species and Elemental Abundances relative to Total Hydrogen}
\label{elem_abund}
\begin{tabular}{c c c c}
\tableline\tableline
Species & Set 1\tnote{[a]} &  Set 2 \tnote{[b]} & Set 3\tnote{[c]}\\
\hline\hline
H$_{2}$ &  5.00$\times$$10^{-1}$ & 5.00$\times$$10^{-1}$& 5.00$\times$$10^{-1}$\\
He &  1.40$\times$$10^{-1}$ & 9.00$\times$$10^{-2}$&  1.40$\times$$10^{-1}$ \\
N & 2.14$\times$$10^{-5}$ & 7.60$\times$$10^{-5}$ & 2.14$\times$$10^{-5}$ \\
O & 1.76$\times$$10^{-4}$ & 2.56$\times$$10^{-4}$& 6.08$\times$$10^{-5}$ \\
F & 2.00$\times$$10^{-8}$ & 6.68$\times$$10^{-9}$& 2.00$\times$$10^{-8}$\\
C$^+$ & 7.30$\times$$10^{-5}$ & 1.70$\times$$10^{-4}$& 7.30$\times$$10^{-5}$ \\
Si$^+$ & 3.00$\times$$10^{-9}$& 8.00$\times$$10^{-9}$& 3.00$\times$$10^{-9}$\\
S$^+$ & 2.00$\times$$10^{-8}$ & 8.00$\times$$10^{-8}$& 2.00$\times$$10^{-8}$\\
Fe$^+$ & 3.00$\times$$10^{-9}$& 3.00$\times$$10^{-9}$& 3.00$\times$$10^{-9}$\\
Na$^+$ & 3.00$\times$$10^{-9}$& 2.00$\times$$10^{-9}$ & 3.00$\times$$10^{-9}$\\
Mg$^+$ & 3.00$\times$$10^{-9}$& 7.00$\times$$10^{-9}$& 3.00$\times$$10^{-9}$\\
Cl$^+$ & 3.00$\times$$10^{-9}$ & 1.00$\times$$10^{-9}$& 3.00$\times$$10^{-9}$ \\
P$^+$ & 3.00$\times$$10^{-9}$& 2.10$\times$$10^{-10}$& 3.00$\times$$10^{-9}$\\
\tableline
\end{tabular}
 \begin{tablenotes}
\item[a] O-Rich, Low-metal Abundances, $\frac{C}{O}$ = 0.41,\\ \citet{Graedel82}
\item[b]$\zeta$ Oph values for C, N, O, $\frac{C}{O}$ = 0.66, \citet{Wakelam10}; \\\citet{Cardelli93,Meyer98}; \\
\citet{Neufeld05,Graedel82}; \\
\citet{Shalabiea95}
\item[c]$\frac{C}{O}$ = 1.2, \citet{Terzieva98}
\end{tablenotes}
\end{center}
\end{threeparttable}
\end{table}

\begin{table*}
\begin{center}
\begin{threeparttable}
\caption{Reactions added to the reaction network}
\label{added_rxns}
\begin{tabular}{c c c c}
\tableline\tableline
Reaction & Rate Coefficient, $k(T)$ & Reaction Network & Ref.\\
\hline\hline
C + HNC $\rightarrow$ HCN + C & 1.20 $\times$ 10$^{-11}$ ($\frac{T}{300 K})^{-0.5}$ cm$^3$ s$^{-1}$ & KIDA, OSU & 1\\
C + HNC $\rightarrow$ CCN + H & 3.00 $\times$ 10$^{-12}$ ($\frac{T}{300 K})^{-0.5}$ cm$^3$ s$^{-1}$ & KIDA, OSU& 1\\
CN + HNC $\rightarrow$ C$_2$N$_2$ + H & 2.00 $\times$ 10$^{-10}$ cm$^3$ s$^{-1}$& KIDA& 2\\
CN + C$_2$H$_6$ $\rightarrow$ HCN + C$_2$H$_5$ & 2.08 $\times$ 10$^{-11}$ ($\frac{T}{300 K})^{-0.22}$ e$^{\frac{-58 K}{T}}$ cm$^3$ s$^{-1}$& KIDA & 3 \\
CCH + HNC $\rightarrow$ H + HC$_3$N\tnote{a} & 1.75 $\times$ 10$^{-10}$ cm$^3$ s$^{-1}$& KIDA& 2\\
HNC + O $\rightarrow$ CO + NH & 7.64  $\times$ 10$^{-10}$ e$^{\frac{-1125 K}{T}}$cm$^3$ s$^{-1}$& OSU& 4 \\
gHNC + gO $\rightarrow$ gCO + gNH & E$_a$ = 1100 K & OSU& 4\tnote{b}\\
gHNC + gO $\rightarrow$ CO + NH & E$_a$ = 1100 K & OSU& 4\tnote{b}\\
gHNC + gH $\rightarrow$ gHCN + gH & E$_a$ = Gas-Phase Barrier & OSU\\
gHNC + gH $\rightarrow$ HCN + H & E$_a$ = Gas-Phase Barrier & OSU\\
\tableline
\end{tabular}
 \begin{tablenotes}
\item[a]Reaction rate updated in reaction network.
\item[b]Grain-surface barrier based on gas-phase barrier of \citet{Lin92}.
\item\textbf{References.} (1) J.-C. Loison, private communication, 2013; (2) \citet{Hebrard12}; (3)\citet{Sims93}, (4) \citet{Lin92}
\end{tablenotes}
\end{threeparttable}
\end{center}
\end{table*}

\bibliographystyle{aa}

\end{document}